\documentclass[preprint,aps,nofootinbib,
superscriptaddress,12pt]{revtex4-1}
\usepackage{amsthm,amsmath,physics,mathtools,amssymb}

\usepackage{charter}

\usepackage{CJKutf8}

\usepackage{enumitem}

\theoremstyle{definition}

\usepackage{graphicx}
\graphicspath{ {images/} }

\usepackage{tensor}

\DeclarePairedDelimiterX{\inp}[2]{\langle}{\rangle}{#1, #2}

\DeclareMathOperator{\sinc}{sinc}

\usepackage{hyperref}
\hypersetup{
    colorlinks=true,
    linkcolor=blue,
    filecolor=magenta,      
    urlcolor=cyan,
}

\usepackage{cleveref}

\usepackage{xparse}

\NewDocumentCommand\LH{mo}{%
  \IfNoValueTF{#2}
   {\mathcal{L}(\mathcal{H}^{#1})}
   {\mathcal{L}(\mathcal{H}^{#1},\mathcal{H}^{#2})}%
}

\newcommand\id{\leavevmode\hbox{\small1\kern-3.3pt\normalsize1}}

\allowdisplaybreaks

\begin{document}

\begin{CJK*}{UTF8}{gbsn}
\title{World quantum gravity: Quantum test objects and Synge's world function}
\author{Ding Jia (贾丁)}
\email{ding.jia@uwaterloo.ca}
\affiliation{Department of Applied Mathematics, University of Waterloo, Waterloo, Ontario, N2L 3G1, Canada}
\affiliation{Perimeter Institute for Theoretical Physics, Waterloo, Ontario, N2L 2Y5, Canada}
\affiliation{Department of Physics and Astronomy, University of Waterloo, Waterloo, Ontario, N2L 3G1, Canada}
\begin{abstract}


A new path integral approach of quantum gravity based on relational variables and quantum test objects is presented. We take as a basic variables the squared invariant distance. This invariant quantity could be technically simpler to work with than variant quantities such as the metric tensor. It also could facilitate the studies of matter coupling and quantum spacetime causal structures. In contrast to approaches based on piecewise linear geometries, here gravity is captured by its effects on quantum test particles and fields. 
By an observation of Parker, under a Feynman sum a gravitational phase can be traded into a Van Vleck-Morette determinant term. 
This leads to a new candidate path integral for gravity, which in certain special cases can be computed efficient in the Lorentzian signature.
We discuss some ambiguities left in the path integral measure, which invite further clarifications.
\end{abstract}
\maketitle
\end{CJK*}

\section{Introduction}

A path integral for quantum gravity coupled to matter takes the schematic form
\begin{align}\label{eq:mf}
A=&\sum_g A_{G}[g] \sum_{\gamma} A_M[\gamma,g],
\end{align}
where $A_{G}[g]$ is the gravitational amplitude for the spacetime configuration $g$, and $A_M[\gamma,g]$ is the matter amplitude for the matter configuration $\gamma$ on the spacetime configuration $g$.
For example, in ordinary quantum field theory,
\begin{align}\label{eq:tmspic}
A=\int \mathcal{D}g_{ab} e^{iS_{G}[g_{ab}]} \int \mathcal{D}\phi e^{iS_M[\phi,g_{ab}]},
\end{align}
where $g$ and $\gamma$ are taken to be the metric and matter fields $g_{ab}$ and $\phi$, and $A_{G}[g]$ and $A_M[\gamma,g]$ are defined using the gravitational and matter actions $S_{G}[g_{ab}]$ and $S_M[\phi,g_{ab}]$.
Alternative non-perturbative approaches 
include Quantum Regge Calculus \cite{Loll1998DiscreteDimensions, Hamber2009QuantumApproach, Barrett2019TullioGravity}, 
(Causal) Dynamical Triangulation \cite{Weingarten1982EuclideanLattice, Loll1998DiscreteDimensions, Ambjrn2001DynamicallyGravity, Ambjorn2012NonperturbativeGravity},
Spin-foam Models and Group Field Theories \cite{Ponzano1968SemiclassicalCoefficients, 
ReisenbergerWorldsheetGravity, Baez1997SpinModels, Perez2013TheGravity, Rovelli2014CovariantGravity, Freidel2005GroupOverview}, 
and Causal Sets \cite{bombelli_space-time_1987, Benincasa2010ScalarSet, Surya2019TheGravity}. 
For additional references on different approaches, see \cite{oriti2009approaches}.

We present a new ``World Quantum Gravity'' (WQG) approach\footnote{Connections to and differences from related approaches is discussed in \cref{sec:cra}.} that takes the squared invariant distance $\sigma(x,y)$ as a basic variable. The approach is so named because $\sigma(x,y)$ is commonly referred to as the ``world function'' \cite{Synge1971Relativity:Theory}. With the affine parametrized $z^a(l)$ as the geodesic connecting $x$ to $y$,
\begin{align}\label{eq:swf}
\sigma(x,y):=\frac{1}{2}(l_y-l_x)\int_{l_x}^{l_y}g_{ab}(z)\frac{dz^a}{dl}\frac{dz^b}{dl} dl.
\end{align}
In words, $\sigma(x,y)$ equals one half the squared geodesic distance between $x$ and $y$.
There are several motivations for choosing $\sigma$ as a basic variable:
\begin{itemize}
\item $\sigma$ is simple. In particular, it is an invariant quantity.
\item $\sigma$ is matter-friendly. Scalar and fermionic fields/particles can be described by worldline configurations and $\sigma$ values along the worldlines. 
\item $\sigma$ indicates causal structure. The relations $\sigma<,=,>0$ manifestly correspond to time-, light-, and space-separations, which facilitates the study of quantum causal structures \cite{HardyProbabilityGravity, *hardy2007towards, chiribella2013quantum, *chiribella2009quantum, oreshkov2012quantum}.
\end{itemize}

In (\ref{eq:swf}) $\sigma$ is derived from $g_{ab}$. However, it is possible to invert the status and describe gravity fundamentally by $\sigma$, because \cite{Synge1971Relativity:Theory}
\begin{align}
g_{ab}(x)=-\lim_{y\rightarrow x}\frac{\partial}{\partial x^a}\frac{\partial}{\partial y^b}\sigma(x,y),
\end{align}
showing that $g_{ab}$ can be recovered from $\sigma$. In principle, one can plug the above expression in (\ref{eq:tmspic}) to change $g_{ab}$ for $\sigma$ to define the theory. However, the result can be unmanageably complicated. 

In this paper we follow an alternative strategy and take a fundamentally relational understanding of gravity.
While the natural home for locating field variables is manifolds, the  natural home for locating relational variables is graphs. We start by assigning $\sigma$ values to edges of graphs, which are taken as elementary structures not embedded in continuum manifolds (\cref{sec:ts}).
The next task is to assign gravitational amplitudes in terms of $\sigma$ on graphs. Previous approaches such as Quantum Regge Calculus and Dynamical Triangulation capture gravity geometrically by curvatures measured on piecewise flat geometries. In contrast, we capture gravity \textit{correlationally} by its effects on the propagation of quantum test fields/particles (\cref{sec:ma} and \cref{sec:ga}).\footnote{A similar strategy is employed in Causal Sets by defining the Benincasa-Dowker gravitational action through introducing scalar matter fields \cite{Benincasa2010ScalarSet, Surya2019TheGravity}. There is also related work on measuring spacetime distances by field correlations \cite{saravani2016spacetime, *KempfReplacingCorrelation}.} 

In describing gravity's influence on classical test objects, the affine connection and the curvature tensors arise as important quantities. For instance, the affine connection appears in the geodesic equation to describe the motion of a classical test particle, and the Riemann curvature tensor appears in the Raychaudhuri equation to describe the relative motion of a congruence of classical test particles. In describing gravity's influence on \textit{quantum} test objects, the Van Vleck-Morette determinant \cite{VanVleck1928TheMechanics., Morette1951OnIntegrals} arises as an important quantity. From the perspective of the path integral, quantum test particles travel along all trajectories, including those that do not obey the classical equation of motion. Gravity's influence is therefore not on the shape of the quantum particle's trajectories, but on its quantum transition amplitudes. It turns out that the Van Vleck-Morette determinant quantifies this influence. Like the affine connection and the curvature tensors, the Van Vleck-Morette determinant which quantifies gravity's influence on test objects can in turn be used to describe gravity itself. This offers a novel approach to quantum gravity.
The particular form of the gravitational amplitude we use derives from a correspondence discovered by Parker \cite{Parker1979PathSpace, Bekenstein1981Path-integralSpacetimeb} that relates the exponential Ricci scalar term to the Van Vleck-Morette determinant.
The path integration formula is fixed after introducing an auxiliary variable $\rho$ which intuitively describes curvature (\cref{sec:pi}). Finally, physical relativistic scalar matter is incorporated through the worldline formulation (\cref{sec:s}).

The models obtained have several merits and several open problems (\cref{sec:c}). By design, the basic variables $\sigma$ and $\rho$ are scalar quantities easy to work with. Relativistic scalar matter can be incorporated straightforwardly, and superposition of spacetime causal structure can be studied at a detailed level. Serendipitously, integration over causal structures enables Monte Carlo simulation when certain conditions are fulfilled. This opens the opportunity to efficiently compute the gravitational path integral for Lorentzian quantum spacetime with many degrees of freedom. The major open problems lie in the ambiguities in the path integral measure regarding integration variables and schemes, as well as possible additional terms. Ideas on overcoming the ambiguities are discussed in \cref{sec:c}.

\section{Locating relational degrees of freedom}\label{sec:ts}


In contrast to $g_{ab}(x)$ as a field variable, $\sigma(x,y)$ is a \textit{relational} variable defined for pairs of locations. 
Whereas the natural arena for the pointwise defined $g_{ab}$ is a manifold, that for the 
pairwise defined $\sigma$ is a 
graph, with locations modelled as points, and neighboring points connected by edges. Specifying a spacetime configuration amounts to assigning a 
value $\sigma_k$ for each edges $k$. 


For concreteness we focus on spacetimes with $1$ temporal and $d=3$ spatial dimensions, and work with $D=4$ dimensional hypercubic lattice graphs. Generalizations to other dimensions and graphs are possible, although we will not investigate these possibilities here.

In this relational approach, spacetime is described fundamentally by graphs with relational gravitational degrees of freedom located on them. The graphs are taken as a primitive structure, not embedded in any continuum spacetime manifolds. Although the graphs are discrete structures, this in no way assumes spacetime is fundamentally discrete. In the gravitational path integral we will integrate over continuous values of spacetime distances on the edges. 



The path integral sum can possibly contain a sum over different graphs. This is needed if quantum gravity involves a sum over different spacetime topologies, or if we adopt the ``summing scheme'' for the path integral discussed in \cref{sec:pi}.



\section{Matter}\label{sec:ma}

Our strategy is to capture gravity through its effects on test relativistic quantum scalar fields/particles. As we review in this section, a quantum scalar field propagator can be re-expressed as a path integral over quantum particle trajectories \cite{Feynman1950MathematicalInteraction, *Feynman1951AnElectrodynamics, Polyakov1987GaugeStrings, Bern1991EfficientAmplitudes, Strassler1992FieldActionsb, SchmidtTheGraphs}. Therefore a test field and particle can be treated in the same way. Later on we will build on this re-expression to not only introduce the amplitude $A_G$ for quantum gravity, but also the amplitude $A_M$ for real (as opposed to test) quantum matter \footnote{This ``worldline method'' to incorporate matter to quantum gravity has been applied before, e.g., in \cite{Freidel2004PonzanoReggeParticles, *FreidelPonzano-ReggeChern-Simons, *Freidel2006PonzanoReggeTheory, Baratin2007HiddenDiagrams, *Baratin2007HiddenFoams}.}.


First consider a scalar field $\phi$ on curved $3+1$-dimension continuum spacetime governed by $(\square+m^2+\xi R)\phi(x)=0$, where $m$ is the mass parameter, $\xi$ is the coupling constant, and $R$ is the Ricci scalar. In the Schwinger proper time representation \cite{Schwinger1951OnPolarization} the Feynman propagator can be expressed as a path integral over trajectories \cite{Parker1979PathSpace, Bekenstein1981Path-integralSpacetimeb}
\begin{align}\label{eq:spt}
G(x,y)=&i\int_0^\infty \braket{x,l}{y,0} e^{-im^2 l} dl,
\\
\label{eq:wlpi}
\braket{x,l}{y,0}=&\int d[x(l')] \exp{i\int_0^l dl' [\frac{1}{4}g_{ab}\frac{dx^a}{dl'}\frac{dx^b}{dl'}-(\xi-\frac{1}{3})R(l')]}
\\
:=&\lim_{N\rightarrow \infty}\Big[\frac{1}{i}(\frac{1}{4\pi i \epsilon})^2\Big]^{N+1}\int \prod_{n=1}^N d^4 x_n [-g(x_n)]^{1/2} 
\nonumber\\
& \exp{\sum_{j=0}^N i\int_{j\epsilon}^{(j+1)\epsilon}[\frac{1}{4}g_{ab}\frac{dx^a}{dl'}\frac{dx^b}{dl'}-(\xi-\frac{1}{3})R(l')] dl' },\label{eq:mac}
\end{align}
with fixed starting point $x_0=x$ and ending point $x_{N+1}=y$. The term $m^2$ is assumed to contain an infinitesimal negative imaginary part in accordance with the Feynman prescription. The variable $l'$ is the Schwinger proper time, and the $l'$ integrals are evaluated along geodesics connecting $x_j$ and $x_{j+1}$. If there are multiple geodesics the shortest is used. The parameter $l$ is partitioned equally to the $N+1$ segments so that $\epsilon:=l/(N+1)$. 


On a graph, the sum over paths is realized as a sum over graph paths. On an edge $k$, the integrand exponential of (\ref{eq:mac}) becomes
\begin{align}\label{eq:ema}
\exp{i\frac{\sigma_k}{2l_k}-i(\xi-\frac{1}{3})R_k l_k},
\end{align}
where $\sigma_k$ and $l_k$ are the squared distance and parameter on edge $k$ (cf. (\ref{eq:swf})). This matter edge amplitude forms the basis for the gravity amplitude we consider next.


\section{Gravity}\label{sec:ga}

Our expression for $A_{G}[\sigma]$ in terms of $\sigma$ is inspired by Parker's remarkable observation \cite{Parker1979PathSpace}. On any spacetime equipped with a metric field, the following exchange is possible under a sum over paths for arbitrary constants $a,b,c$ (see \Cref{sec:po} for an explanation),
\begin{align}\label{eq:pmf1}
\Delta^a \exp{i[\frac{\sigma}{2l}-(\frac{a}{3}-c)Rl]} \xleftrightarrow{\sum_\text{path}} \Delta^b \exp{i[\frac{\sigma}{2l}-(\frac{b}{3}-c)Rl]}.
\end{align}
Here $\xleftrightarrow{\sum_\text{path}}$ indicates that the correspondence between the left and right hand sides holds only inside a path integral.
The quantity $\Delta$ is the Van Vleck-Morette determinant \cite{VanVleck1928TheMechanics., Morette1951OnIntegrals, Visser1993VanSpacetimes} (see \cref{sec:vd} for an introduction to this important quantity), a functional of $\sigma$ and its second order derivative. Explicitly,
\begin{align}\label{eq:vdd}
\Delta(x,y)=C s^d \exp{-\int \theta ds'},
\end{align}
where $d=3$ is the spatial dimension, $s=\abs{2\sigma(x,y)}^{1/2}$, $\theta(x)=(\tensor{\sigma}{^a_a}(x,y)-1)/\abs{2\sigma(x,y)}^{1/2}$, $\tensor{\sigma}{^a_a}(x,y)$ is the trace of the matrix $\tensor{\sigma}{^a_b}(x,y)=\nabla_{x^b}\nabla^{x^a}\sigma(x,y)$, the integral is along the geodesic from $y$ to $x$, and $C$ is a constant determined by the boundary condition $\Delta(y,y)=1$. 
Later we will use a particular case of (\ref{eq:pmf1}) for $a=0$ and $b=3c$, 
\begin{align}\label{eq:pmf2}
\exp{i(\frac{\sigma}{2l}+c Rl)} \xleftrightarrow{\sum_\text{path}} \Delta^{3c} \exp{i\frac{\sigma}{2l}}.
\end{align}
This means the $R$ term in (\ref{eq:ema}) can be traded into a $\Delta$ term.

Importantly, Parker's result holds on any spacetime equipped with a metric field, which need not solve the classical equation of motion of GR (see \Cref{sec:po}). This means the exchange can be applied to spacetime configurations under the path integral, even though they do not solve the classical equation of motion.

To obtain the gravitational amplitude $A_{G}[\sigma]$, the plan is to extrapolate the above correspondence to the graph setting to trade away an analogue Einstein-Hilbert action term for an expression in terms of $\sigma$. 

Motivated by the form of the Einstein-Hilbert action, a $D$ dimensional elementary lattice cube $\kappa$ and edge $h$ in it is initially formally assigned a gravitational amplitude $\exp{i \bar{\alpha} d^4x_{\kappa,h} \sqrt{-g_h} R_h}$, which combines into $\prod_{h\in\kappa}\exp{i \bar{\alpha} d^4x_{\kappa,h} \sqrt{-g_h} R_h}$ to become the gravitational amplitude of the cube $\kappa$, which in turn combines into $\prod_{\kappa\subset \Gamma}\prod_{h\in\kappa}\exp{i \bar{\alpha} d^4x_{\kappa,h} \sqrt{-g_h} R_h}$ to become the gravitational amplitude of the graph $\Gamma$. Here  $\bar{\alpha}$ is a coupling constant, $d^4x_{\kappa,h} \sqrt{-g_h}$ is the spacetime volume attributed to $(\kappa,h)$, and $R_h$ is the Ricci curvature attributed to $h$. First we want to find an expression for $d^4x_{\kappa,h} \sqrt{-g_h}$ in terms of $\sigma$. We know that on a manifold in a Riemann normal coordinate system around $x$, $\sqrt{-g(y)}\approx\Delta^{-1}(x,y)$ \cite{Poisson2011}. This motivates the prescription $\sqrt{-g_h}\rightarrow \Delta_h^{-1}$ where $\Delta_h$ is the Van Vleck-Morette determinant evaluated on $h$. 
For $d^4x_{\kappa,h}$, let 
\begin{align}
s_i:=\abs{2\sigma_i}^{1/2}
\end{align}
be the physical proper distance along any edge $i$, and define the average expression
\begin{align}\label{eq:qgaf}
\eta_{\kappa,h}=& \alpha \sum_{C_{\kappa,h}}\frac{1}{2}\prod_{h\in C_{\kappa,h}} s_{h}
\end{align}
where $\alpha$ is a constant parameter, and the average is over the two corners $C_{\kappa,h}$ containing the edge $h$ within the cube $\kappa$. In $D$ spacetime dimensions, each set of $D$ edges belonging to the same lattice cube and sharing a same vertex forms a corner.
This leads to the prescription $\bar{\alpha} d^4x_{\kappa,h} \rightarrow \eta_{\kappa,h}$.

This is a \textit{prescription}, because the graph is not assumed to be embedded in a manifold, so there is not a uniquely preferred way to define spacetime volumes. One could reason in a Wilsonian fashion that there is some broad theory space for different quantum amplitudes as functions of $\sigma$, and that additional terms could be included in (\ref{eq:qgaf}). In this context, (\ref{eq:qgaf}) as it stands is our guess for the most significant term, and additional terms are to be included in the measure factor of \Cref{sec:pi}.
By not equating $\alpha$ and $\bar{\alpha}$, we bring in an additional free parameter $c:= \alpha/\bar{\alpha}$ to the prescription to give it more flexibility. The spacetime volume attributed to $(\kappa,h)$ is then $V_{\kappa,h}:=c \Delta^{-1}_h \eta_{\kappa,h}$. The values $\alpha$ and $c$ are ultimately determined by matching with experimental data as in renormalization of QFT. 


So far we arrived at the gravitational amplitude $\exp{i\eta_{\kappa,h} \Delta^{-1}_h R_h}$ for the cube-edge pair $(\kappa,h)$.
We now apply (\ref{eq:pmf2}) to obtain an expression completely in terms of $\sigma$. 
\begin{align}\label{eq:dsp}
\exp{i\eta_{\kappa,h} \Delta_h^{-1} R_h}
=& G_h^{-1} G_h \exp{i\eta_{\kappa,h} \Delta_h^{-1} R_h}
\\
=&G_h^{-1} \int_0^\infty \frac{dl_h}{(4\pi i l_h)^{2}}\exp{i\frac{\sigma_h}{2l_h}}\exp{i\frac{\eta_{\kappa,h} \Delta_h^{-1}}{l_h} R_h l_h}\label{eq:dsp2}
\\
\rightarrow& G_h^{-1} \int_0^\infty \frac{dl_h}{(4\pi i l_h)^{2}}
\Delta_h^{3 \eta_{\kappa,h} \Delta_h^{-1}/ l_h}\exp{i\frac{\sigma_h}{2l_h}}\label{eq:dsp3}
\\
=& \frac{\sigma_h}{\sigma_h-6i \eta_{\kappa,h} \Delta_h^{-1} \ln \Delta_h}.
\label{eq:dep}
\end{align}
We introduced a factor of $1=G_h^{-1} G_h$, with $G_h:=\int_0^\infty \frac{dl_h}{(4\pi i l_h)^{2}}\exp{i\frac{\sigma_h}{2l_h}}$ as the Feynman propagator for a massless test particle ($\sigma_h$ implicitly contains an infinitesimal imaginary part $i\epsilon$ to make the integral converge). In the second and third steps we extracted a factor of $l_h$ in the gravitational exponent, and applied (\ref{eq:pmf2}). 
Since (\ref{eq:pmf2}) holds under a path sum, here it is applied to an 1-segment approximation of the path integral (a special case of the general $N+1$-segment approximation in (\ref{eq:mac})). One can conceive increasing the number of segments to improve the approximation. However, this is most naturally done by moving to a graph with more edges and vertices \footnote{Suppose one increases the number of segments on each edge by breaking it into multiple edges. In the present setting each new edge should be associated with a new $\sigma$, and the natural way to conduct the path integral is to sum over $\sigma$ on the new edges. An even larger regular lattice graph which contains this enlarged one as a subgraph further improves the approximation. In the end the improvement of the approximation can be conducted on a larger regular graph with an 1-segment approximation.}. In the end, still a 1-segment approximation is used on each edge.
In the last step we applied $\int_0^\infty \frac{dl}{l^2}\exp{i\frac{X+i\epsilon}{l}} = i/(X+i\epsilon)$ (easily verified by changing variable from $l$ to $1/l$) to the $l_h$ integral and the integral in $G_h^{-1}$. 

Above we used a massless test particle. If a massive test particle with the propagator $G_h:=\int_0^\infty \frac{dl_h}{(4\pi i l_h)^{2}}\exp{i\frac{\sigma_h}{2l_h}-im^2 l_h}$ is used, we get in place of (\ref{eq:dep})
\begin{align}\label{eq:dsp1}
\exp{i\eta_{\kappa,h} \Delta_h^{-1} R_h}
\rightarrow & G_h^{-1} \int_0^\infty \frac{dl_h}{(4\pi i l_h)^{2}}
\Delta_h^{3 \eta_{\kappa,h} \Delta_h^{-1}/ l_h}\exp{i\frac{\sigma_h}{2l_h}-im^2 l_h}.
\end{align}
The expression (\ref{eq:dep}) appears simpler, so we will work with it in the following. However, other than simplicity we do not know any reason that disfavours (\ref{eq:dsp1}), which seems to be a feasible prescription. This leaves an one-parameter ambiguity parameterized by the test particle mass.

\section{Path integral}\label{sec:pi}

Given (\ref{eq:dep}) as the gravitational amplitude on a cube-edge pair, the total gravitational amplitude on a graph $\Gamma$ should be
\begin{align}\label{eq:gga}
\prod_{\kappa\subset \Gamma}\prod_{h\in\kappa} \frac{\sigma_h}{\sigma_h-6i \eta_{\kappa,h} \Delta_h^{-1} \ln \Delta_h},
\end{align}
where the product is over all elementary lattice cubes $\kappa$ of $\Gamma$ and over all edges $h$ of fixed $\kappa$.
To fix the gravitational path integral, we need to discuss the summation procedure. Specifically, we need to fix the integration variables and their integration ranges, and discuss whether to sum over graphs.

The graph amplitude (\ref{eq:gga}) depends on $\Delta$, which as given in (\ref{eq:vdd}) depends on both $\sigma$ itself (through $s=\abs{2\sigma}^{1/2}$) and its second order derivative $\tensor{\sigma}{^a_a}$ (through $\theta=[\tensor{\sigma}{^a_a}-1]/s$). One strategy for performing the path integral is to introduce an auxiliary variable $\rho$ independent of $\sigma$ to replace the second order derivative, and then integrate over both $\sigma$ and $\rho$. This is analogous to the strategy of the first order formalism \cite{Palatini1919DeduzioneHamilton} that treats the affine connection as a variable independent of metric.

We want to use the Raychaudhuri equation to introduce the auxiliary variable and obtain an alternative expression for $\theta(s)$ to be used in (\ref{eq:vdd}). Recall the Raychaudhuri equation for timelike \cite{Poisson2004AToolkit} and spacelike \cite{Abreu2011SomeEquation} geodesic congruences (In the path integration over $\sigma$ specified below in (\ref{eq:pim}), lightlike separation $\sigma=0$ is of measure zero and hence is unimportant.):
\begin{align}\label{eq:re}
\frac{d\theta}{ds}=-\frac{1}{3}\theta^2-\bar{\sigma}^2+\omega^2-R_{ab}u^a u^b.
\end{align}
Here $s$ is an affine parameter along the geodesics, $\theta$ is the expansion, $\bar{\sigma}^2$ is the squared shear, $\omega^2$ is the squared rotation, $R_{ab}$ is the Ricci tensor, and $u^a$ is the unit tangent vector along the geodesics.
We assume that
\begin{align}
\rho=\bar{\sigma}^2-\omega^2+R_{ab}u^a u^b
\end{align} 
is constant on each edge. This is similar in spirit to the standard prescription of piecewise linear (vanishing acceleration and constant velocity) trajectories in evaluating the path integral for a point particle \cite{Feynman1965QuantumIntegrals}. Intuitively, $\rho$ quantifies spacetime curvature, which vanishes for a flat spacetime where $\bar{\sigma}=\omega=R_{ab}u^a u^b=0$.
Equation (\ref{eq:re}) then becomes $\frac{d\theta}{ds}=-\frac{1}{3}\theta^2-\rho$. The quantity $\rho$ controls the rate of change of the expansion $\theta$. The larger $\rho$ is, the faster geodesic congruence shrinks. From the coincidence limit of $\tensor{\sigma}{_a_b}(x,y)=\nabla_{x^b}\nabla_{x^a}\sigma(x,y)$, one can derive the boundary condition $\theta(0)=\infty$ \cite{Poisson2011}. The solution to the differential equation is then
\begin{align}\label{eq:tua}
\theta(s)=\sqrt{3\rho}\cot(s\sqrt{\frac{\rho}{3}}).
\end{align}
One can check that in the flat spacetime limit $\rho\rightarrow 0$, the familiar expression $\theta(s)\rightarrow 3/s$ is recovered.

Plugging (\ref{eq:tua}) in (\ref{eq:vdd}) and using the boundary condition $\Delta(0)=1$, we obtain
\begin{align}\label{eq:vds}
\Delta(s,\rho)=\bigg[s\sqrt{\frac{\rho}{3}} ~\csc(s\sqrt{\frac{\rho}{3}})\bigg]^3=\sinc^{-3} \bigg(s\sqrt{\frac{\rho}{3}}\bigg).
\end{align}
When this expression is used, (\ref{eq:gga}) 
no longer depends on the second order derivative $\tensor{\sigma}{^a_a}$. All the spacetime degrees of freedom are now encoded in the values of $\sigma_h$ and $\rho_h$ assigned to the edges $h$.

Denoting the integration part of the path integral by the symbol $\sum_g$, we specify the integration range as
\begin{align}\label{eq:pim}
\sum_g=&\prod_{j\in\Gamma} \int_{-\infty}^{\infty} d\rho_j \int_{-a(\rho_j)^2/2}^{a(\rho_j)^2/2} d\sigma_j,
\\
a(\rho)=&
\begin{cases}
b \pi \sqrt{\frac{3}{\rho}}, \quad \rho>0,
\\
\infty, \quad \rho\le 0,
\end{cases}
\end{align}
with $b$ obeying $0<b<1$ as a parameter for this evaluation scheme.
To explain this prescription, we first changing variable from $\sigma=\pm s^2/2$ to $s$. For $\rho>0$,  by (\ref{eq:tua}) we have $\theta\rightarrow -\infty$ as $s\rightarrow \pi\sqrt{3/\rho}$. In the continuum, at this caustic point where the expansion parameter blows up, $\sigma$ stops being well-defined as there are multiple geodesics connecting the two points. In the graph setting, this suggests the introduction of a new vertex before reaching this point. Practically this amounts to integrating $s$ only up to $a = b \pi(3/\rho)^{1/2} < \pi(3/\rho)^{1/2}$ for some $0<b<1$. Equivalently, we integrate $\sigma$ from $-a^2/2$ to $a^2/2$. For $\rho\le 0$, there is no divergence for $\theta$ according to  (\ref{eq:tua}). Therefore for $\rho\le 0$ we integrate $\sigma$ from $-\infty$ to $\infty$. The integration range of $\rho$ is taken from $-\infty$ to $\infty$.

Now we come to the question whether to sum over graphs or not. In a refinement scheme, the exact result is approached by going to larger graphs with more degrees of freedom. In a summing scheme, the exact result is approached by summing over more graphs. It is possible that different schemes lead to the same result if suitable path integral measure factors are assigned. This is the case for the path integral of a point particle (\cref{sec:pis}), and also for certain spin-foam models of quantum gravity \cite{Rovelli2012InRefining}. Without knowing any reason to rule out either scheme, we keep both schemes under consideration for the current models.\footnote{The above discussion is assuming that spacetime topology is fixed. In situations where summing over spacetime topology is needed, in both schemes we should include an additional sum over graphs representing different topologies.}

To sum up, the path integral for pure gravity has the expression
\begin{align}\label{eq:pga1}
A_{G}=\lim_{\Gamma}\sum_g \mu_{r}(\sigma,\rho,\Gamma) \prod_{\kappa\subset \Gamma}\prod_{h\in\kappa} \frac{\sigma_h}{\sigma_h-6i \eta_{\kappa,h} \Delta_h^{-1} \ln \Delta_h}
\end{align}
in a graph refinement scheme, and
\begin{align}\label{eq:pga2}
A_{G}=\sum_{\Gamma} \sum_g \mu_{s}(\sigma,\rho,\Gamma) \prod_{\kappa\subset \Gamma}\prod_{h\in\kappa} \frac{\sigma_h}{\sigma_h-6i \eta_{\kappa,h} \Delta_h^{-1} \ln \Delta_h}
\end{align}
in a graph summing scheme. In the refinement scheme $\lim_\Gamma$ approaches ever larger graphs with more edges and vertices, 
while in the summing scheme, $\sum_{\Gamma}$ sums over graphs. At this stage, it is not clear if it is necessary to include graphs beyond regular lattices in the limit or the sum. One possibility is that through further study we can identify universal behaviors that renders other graphs unnecessary. In the expressions for the amplitudes, $\sigma$ and $\rho$ are the variables representing squared spacetime distance and curvature. On an edge $h$ they take values $\sigma_h$ and $\rho_h$. The variable $s=\abs{2\sigma}^{1/2}$ is the geodesic distance, and the Van Vleck-Morette determinant $\Delta$ in terms of $s$ and $\rho$ is given in (\ref{eq:vds}). The product $\prod_{\kappa\subset \Gamma}\prod_{h\in\kappa}$ is over all elementary lattice cubes $\kappa$ of $\Gamma$ and over all edges $h$ of $\kappa$. The different measure factors $\mu_{r}$ and $\mu_{s}$ are introduced to account for the different refinement and summing schemes. For suitable assignment of the factors the two schemes could give the same results, in which case the choice of scheme is up to our convenience. The measure factors, like Wilsonian actions of QFTs, are generic functions that hold our uncertainty about the exact form of the path integral. Some ideas on reducing the uncertainties are discussed in \cref{sec:c}. The integration $\sum_g$ is specified in (\ref{eq:pim}).\footnote{Because we integrate over continuous values, configurations equivalent up to graph relabelling are of measure zero, and this kind of redundancy seems harmless. In any case one can introduce a symmetry factor in the measure factor in case it is relevant.} Like any path integral, suitable boundary conditions/constraints should be imposed in practical applications. 


Suppose the measure factors $\mu(\sigma,\rho,\Gamma)$ of (\ref{eq:pga1}) and (\ref{eq:pga2}) are cube-wise local, i.e., decomposable into a cube-wise products as in $\mu(\sigma,\rho,\Gamma)=\prod_{\kappa\subset \Gamma} \nu_\kappa(\sigma_\kappa,\rho_\kappa)$ where $\sigma_\kappa,\rho_\kappa$ are data in cube $\kappa$. Then the path integral integrands are products of cube-wise contributions. In this case composition of multiple regions is straightforward. One simply multiplies the cube factors from different regions to obtain the integrand of the composite region. In this case, no additional boundary term is needed to ensure that the path integral behave well under composition.

\section{Coupling matter}\label{sec:s}

In \cref{sec:ma}, we reviewed the path integral expression for the Feynman propagator of a scalar field coupled to gravity. This offers a way to couple scalar matter to the model of quantum gravity. 

On a manifold the matter amplitude can be expressed as a sum over Feynman diagrams, where the diagram edges represent the Feynman propagators. In terms of the path integral expression (\ref{eq:spt}) for the propagator, the sum over Feynman diagrams becomes a sum over \textit{correlation diagrams} \cite{Jia2021CorrelationalConstraints}, which are Feynman diagrams whose edges are replaced by curves localized in spacetime. A curve bounded by $x$ and $y$ contributes as an amplitude the integrand of (\ref{eq:wlpi}) evaluated along the curve. The sum over curves sharing the same vertices amounts to the path integral of (\ref{eq:wlpi}), supplemented by the residue integral of (\ref{eq:spt}), eliminating the dependence on the artificial parameter $l$. 
Intuitively, the correlation diagrams with localized curves and vertices represent ways matter correlations are mediated in spacetime, and provides a ``basis'' of matter configurations $\gamma$ in (\ref{eq:mf}).

\begin{figure}
    \centering
    \includegraphics[width=.7\textwidth]{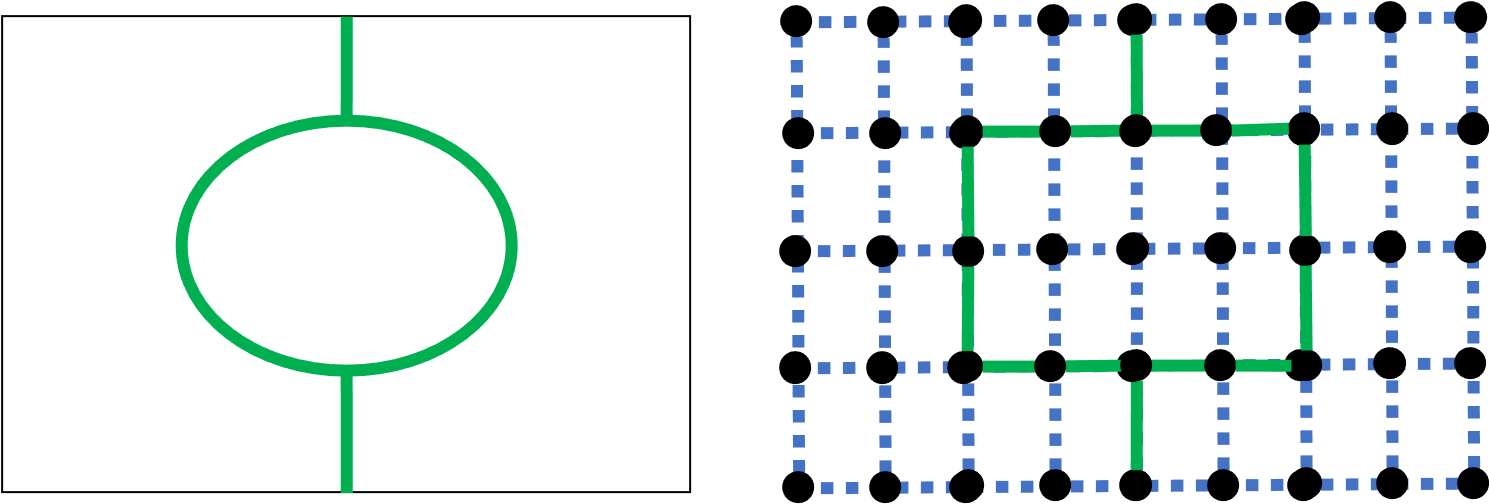}
    \caption{A matter Feynman diagram in ordinary QFT (left) and one of its corresponding correlation diagrams in World Quantum Gravity (right). The correlation diagram is schematically shown on a $2D$ spacetime lattice graph $\Gamma$, consisting of dashed pure gravity edges, and solid matter edges forming a subgraph $\gamma$.}
    \label{fig:wqgpic}
\end{figure}

On a graph $\Gamma$ of the quantum gravity model, a correlation diagram becomes a subgraph $\gamma$, as illustrated in \Cref{fig:wqgpic}. Each edge $k\in \gamma$ is assigned a matter amplitude $[i(4\pi i l_k)^{2}]^{-1}\exp{i\frac{\sigma_k}{2l_k}-i(\xi-\frac{1}{3})R_k l_k}$, an analogue (cf. (\ref{eq:ema})) of the segment amplitude in (\ref{eq:mac}). 
$l_k$ is an unphysical parameter that the physical amplitude should not depend on. On a manifold, $l$ is eliminated by the integral in (\ref{eq:spt}). The $l$-integral is from vertex to vertex, but does not extend beyond vertices. The fundamental reason is that the reparametrization invariance of the curves do not extend beyond vertices that connect three or more curves (shifting such vertices along the curve changes the physical configuration). 
On a graph there is an ambiguity. If the gravitational edges are thought of as analogues of gravitational propagators, then the matter $l$-integral should be performed on each edge. If the gravitational edges are thought of as analogues of edges in a lattice field theory, the matter $l$-integral seems more reasonably performed on each path of $\gamma$. 

We study the first case here, and the second case in \cref{sec:acsm}. With the $l$-integrals performed on each matter edge, the matter amplitude is
\begin{align}\label{eq:ma1}
A_M[\gamma,g]=&V[\gamma]\prod_{k\in\gamma}A_M[k,g],
\\
A_M[k,g]=&\int_0^\infty \frac{dl_k}{(4\pi i l_k)^{2}}\exp{i\frac{\sigma_k}{2l_k}-i(\xi-\frac{1}{3})R_k l_k-im^2l_k}.
\label{eq:mae}
\end{align}
The vertex factor $V[\gamma]$ encodes the interaction coupling constants and symmetry factors as in ordinary Feynman diagrams. Like the interaction potential $V$ of a matter QFT, the form of $V[\gamma]$ is not fixed \textit{a priori}, but depends on theory. From a known potential of a matter QFT the corresponding $V[\gamma]$ can be obtained by lattice worldline models reviewed in \cite{Gattringer2016ApproachesTheory}.

Given $\Gamma$ as a spacetime graph, and $\gamma_\Gamma$ as a matter subgraph, we combine $A_M[\gamma_\Gamma,g]$ with the gravitational edge amplitudes to obtain
\begin{align}\label{eq:gmab}
A[\Gamma,\gamma_\Gamma,g] = V[\gamma_\Gamma]\prod_{\kappa\subset \Gamma}\prod_{h\in\kappa} \frac{\sigma_h}{\sigma_h-6i \eta_{\kappa,h} \Delta_h^{-1} \ln \Delta_h} 
\prod_{k\in \gamma_\Gamma}\Delta_k^{1-3\xi} \int_0^\infty \frac{dl_k}{(4\pi i l_k)^{2}}
\exp{i\frac{\sigma_k}{2l_k}-im^2l_k},
\end{align}
where as in the pure gravity case we applied (\ref{eq:pmf2}) to the matter amplitude to trade away the Ricci curvature term for a $\Delta$ term. Bringing in $\sum_g$ defined in (\ref{eq:pim}) leads to the results
\begin{align}\label{eq:gma1}
A_{GM}=\lim_{\Gamma} \sum_{\gamma_\Gamma} \sum_g \mu_{r}(\sigma,\rho,\Gamma) A[\Gamma,\gamma_\Gamma,g]
\end{align}
in a graph refinement scheme, and
\begin{align}\label{eq:gma2}
A_{GM}=\sum_{\Gamma} \sum_{\gamma_\Gamma} \sum_g \mu_{s}(\sigma,\rho,\Gamma) A[\Gamma,\gamma_\Gamma,g]
\end{align}
in a graph summing scheme. Here $\sum_{\gamma_\Gamma}$ are sums over matter subgraphs $\gamma_\Gamma$ of the spacetime graph $\Gamma$. The measure factors $\mu_r$ and $\mu_s$ are the same as in (\ref{eq:pga1}) and (\ref{eq:pga2}). 

\section{Discussion}\label{sec:c}


We have presented a new path integral approach to quantum gravity for Lorentzian spacetimes. 
The main ideas are recapitulated as follows.
\begin{enumerate}
\item Adopt a relational understanding of gravity and choose the invariant squared spacetime distance $\sigma$ (``world function'') as the basic variable.
\item Locate the relational variables on graphs, the canonical mathematical structure for studying relations.
\item Define the gravitational amplitude by gravity's effects on \textit{quantum} test fields/particles.
\item Introduce an auxiliary curvature variable $\rho$ for performing the path integrals.
\end{enumerate}
We regard the first three points as the defining aspects of this approach, which we called ``World Quantum Gravity''. In this approach, discrete graphs are elementary structures that replaces continuum manifolds for locating physical degrees of freedom according to (2), but spacetime distances take continuous values according to (1). This approach does not assume fundamental discreteness for spacetimes in the sense that spacetime distances take continuous values, but it should not be viewed as an alternative regularization prescription for continuum quantum gravity because the elementary structure are discrete graphs rather than continuum manifolds. In addition, the path integral amplitudes are obtained in a different way according (3).
Point (4) is a choice made to fix the variables for the path integration. Models compatible with the first three points but with different choices of variables may exist. Such alternative models are worth exploring, just like field theories of gravity with alternative variables are. 

We find the approach interesting for several reasons. 
The basic variables $\sigma$ and $\rho$ are simple scalar quantities easy to work with. As shown, scalar matter can be incorporated through the worldline formulation. 
The incorporation of the superposition of spacetime causal structures is manifest. On each edge, timelike, lightlike, and spacelike causal relations correspond to $\sigma$ taking negative, zero, and positive values. In the path integral all three kinds of $\sigma$ values are integrated over. In this sense, the vertices of the edge is in a superposition of timelike, lightlike, and spacelike causal relations. Therefore the models can then be used to study quantum indefinite causal structures \cite{HardyProbabilityGravity, *hardy2007towards, chiribella2013quantum, *chiribella2009quantum, oreshkov2012quantum} of spacetime at a detailed level. As discussed below, the integration over causal structures removes the ``sign problem'' for Monte Carlo simulation for certain measure factors and when the path integration is unconstrained \cite{JiaUltravioletStructure}. This opens the opportunity to compute the gravitational path integral for Lorentzian quantum spacetime with many degrees of freedom in certain cases.


In a Wilsonian approach to QFT, the ambiguities in the path integral measure are collected in the action, which \textit{a priori} contains infinitely many terms with different significance. In the models considered here, the ambiguities in the path integral measure are collected in $\mu_r(\sigma,\rho,\Gamma)$ and $\mu_s(\sigma,\rho,\Gamma)$, which are analogues of the Wilsonian action in the graph refining and graph summing schemes, respectively. For a QFT of gravity, the Einstein-Hilbert term is the dominating contribution to the action. Since the edge amplitudes of (\ref{eq:pga1}) and (\ref{eq:pga2}) are obtained from the correspondence to the Einstein-Hilbert action through Parker's formula, the dominating contribution is expected to come from setting $\mu_r(\sigma,\rho,\Gamma)$ and $\mu_s(\sigma,\rho,\Gamma)$ to $1$. 

However, this leaves us with several ambiguities in the path integral measure regarding other contributions from:
\begin{itemize}
\item Integration variables and schemes. The choice of integration variables $(\sigma,\rho)$ may be accompanied by a measure factor (e.g., a Jacobian factor if they come from transforming other variables). In addition, the summing and refining schemes should have different measure factors, as discussed in the paragraph above the paragraph of (\ref{eq:pga1}).
\item Additional terms. For instance, there could be additional terms analogous to higher order terms in QFT actions. There could also be additional terms associated with spacetime volume, as discussed in the paragraphs after (\ref{eq:qgaf}), and an additional term from giving mass to the test particle, as discussed around (\ref{eq:dsp1}).
\end{itemize}
Besides, as discussed in \cref{sec:s} and \cref{sec:acsm}, there exist different ways to couple scalar matter to quantum gravity. 

These ambiguities should come as no surprise. Measure ambiguities are generic among path integral approaches to quantum gravity \cite{Hamber1999OnGravity, Bianchi2010TheGravity, Bahr2010OperatorModels, Dittrich2011PathGravityb, Laiho2011EvidenceGravityb}. 
In ordinary QFT, the ambiguities in the Wilsonian action are reduced by both theoretical (e.g., renormalization group analysis) and experimental (e.g., measurements of coupling constants) considerations. We expect the same for the present models. On the theoretical side, we should look for principles/conditions that constrain the theory space. One example is to require that the path integrals are free from ultraviolet divergences.\footnote{In non-gravitational settings one could also impose unitarity to constrain the path integral measure. For a gravitational path integral without a singled out global time it is not clear whether this requirement is meaningful \cite{Hamber2009QuantumApproach}.} In addition, renormalization group type studies that relate theories on different graphs can also be helpful. On the experimental side, theorists should at least look for efficient ways to compute the path integral in preparation to compare with data.

In \cite{JiaUltravioletStructure} we show that for certain measures the path integral is free of ultraviolet divergences, even if the integration range includes zero invariant distance $\sigma=s=0$. The intuition is that summing over causal structures smears away the divergence \cite{Deser1957, Ford1995GravitonsFluctuations, Ohanian1997, Ohanian1999}. 
Future works could aim at understanding this mechanism for a broader family of measures. 
In \cite{JiaUltravioletStructure} we also note that for certain measures the bulk path integral is free from the Monte Carlo sign problem thanks to the sum over causal relations. 
Recall from (\ref{eq:pga1}) and (\ref{eq:pga2}) that for the pure gravity, the edge amplitude on edge $h$ takes the form 
\begin{align} 
A_h=\frac{\sigma_h}{\sigma_h-6i \eta_{\kappa,h} \Delta_h^{-1} \ln \Delta_h}.
\end{align}
Flipping the sign of $\sigma_h$ and adding $A_h(-\sigma_h)$ to the original $A_h(\sigma_h)$, we have
\begin{align} 
A_h(\sigma_h)+A_h(-\sigma_h)=&\frac{\sigma_h}{\sigma_h-6i \eta_{\kappa,h} \Delta_h^{-1} \ln \Delta_h}+\frac{-\sigma_h}{-\sigma_h-6i \eta_{\kappa,h} \Delta_h^{-1} \ln \Delta_h}
\\
=&\frac{2\sigma_h^2}{\sigma_h^2+(6 \eta_{\kappa,h} \Delta_h^{-1} \ln \Delta_h)^2}\ge 0.
\end{align}
In the first line use noted that flipping the sign of $\sigma_h$ does not change $s_h$ and hence leaves $\eta$ and $\Delta$ invariant according to their definitions (\ref{eq:qgaf}) and (\ref{eq:vds}). Therefore resumming $A_h$ over positive and negative $\sigma_h$ values results in a non-negative number. Provided the measure factor does not destroys this property and is also non-negative, the integral of (\ref{eq:pga1}) and (\ref{eq:pga2}) for the pure gravity is free from the sign problem.

This avoidance of the sign problem is quite delicate. It holds for pure gravity without matter, it requires that $\sigma$ is integrated evenly over positive and negative values on all edges, and it applies to a subclass of the measure factors. Nevertheless when these conditions are fulfilled, Markov Chain Monte Carlo simulation can be applied efficiently to compute the path integral.
The situation is to be contrasted with that of Causal Dynamical Triangulations \cite{Ambjorn2012NonperturbativeGravity}. There Markov Chain Monte Carlo simulation is possible after timelike distances on lattice edges are analytically continued to spatial values to overcome the sign problem. Here no analytic continuation is needed. Both approaches have prices to pay. For Causal Dynamical Triangulations, it is not knowing in general how to analytically continue back \cite{Ambjorn2012NonperturbativeGravity}. For the current approach, it is the above-mentioned limitations.
When these conditions are not fulfilled, more advanced methods are needed. For instance, the Lefschetz thimble method \cite{AlexandruComplexProblem} offers an interesting candidate method to ameliorate the sign problem.

\section*{Acknowledgement}
I am very grateful to Lucien Hardy, Achim Kempf, Laurent Freidel, Rafael Sorkin, Erik Schnetter, and Dustin Lang for valuable discussions at various stages.

Research at Perimeter Institute is supported in part by the Government of Canada through the Department of Innovation, Science and Economic Development Canada and by the Province of Ontario through the Ministry of Economic Development, Job Creation and Trade. This publication was made possible through the support of the grant ``Causal Structure in Quantum Theory'' from the John Templeton Foundation and the grant ``Operationalism, Agency, and Quantum Gravity'' from FQXi. The opinions expressed in this publication are those of the author and do not necessarily reflect the views of the funding agencies. 


\appendix

\section{The world function}

We introduce the world function and collect some relevant formulas in this section. See \cite{Synge1971Relativity:Theory, Poisson2011} for a more comprehensive treatment of the formalism and formulas quoted below.

Consider two points $x,y$ on a spacetime manifold such that $y$ lies in the convex normal neighborhood of $x$. Parametrize the unique geodesic $z^a(l)$ connecting $x$ to $y$ by an affine parameter $l$. Define $\sigma(x,y)$ by
\begin{align}\label{eq:swf1}
\sigma(x,y)=\frac{1}{2}(l_y-l_x)\int_{l_x}^{l_y}g_{ab}(z)\frac{dz^a}{dl}\frac{dz^b}{dl} dl.
\end{align}

This is nothing but one half the squared geodesic distance. For example, on flat spacetime (\ref{eq:swf1}) reduces to the familiar expression $\sigma(x,y)=\frac{1}{2}\eta_{ab}(y-x)^a(y-x)^b$. On a general curved spacetime, first let us introduce 
\begin{align}
t^a=\frac{dz^a}{dl}.
\end{align}
Then $\frac{D}{dl}t^a=0$, since $z^a(l)$ is a geodesic. Hence $c=g_{ab}(l)\frac{dz^a}{dl}\frac{dz^b}{dl}=g_{ab}t^a t^b$ is constant along $z^a(l)$, and $\sigma(x,y)=\frac{c}{2}(l_y-l_x)^2$. Picking $l$ to be the proper distance $s$ implies $c=\pm 1$ for spacelike and timelike separations, whence $\sigma(x,y)=\pm s^2/2$. In the null case $\sigma(x,y)=0=s$. Therefore the formula
\begin{align}
\sigma(x,y)=\pm s^2/2
\end{align}
can be used for all three cases. This equation expresses $\sigma$ in terms of $s$. Conversely, the proper distance $s=\abs{2\sigma}^{1/2}$ can be expressed in terms of $\sigma$.

Synge calls $\sigma(x,y)$ the \textit{world function}, because ``it determines the curved world of space-time'' \cite{Synge1971Relativity:Theory}. Indeed one can show that $\sigma$ knows all about $g_{ab}$:
\begin{align}\label{eq:mwf}
g_{ab}(x)=-\lim_{y\rightarrow x}\frac{\partial}{\partial x^a}\frac{\partial}{\partial y^b}\sigma(x,y).
\end{align}
Here partial instead of covariant derivatives are used because with $x$ held fixed, $\sigma(x,y)$ is a scalar field at $y$, and with $y$ held fixed, $\frac{\partial}{\partial y^b}\sigma(x,y)$ are scalar fields at $x$. 

Higher order covariant derivatives of $\sigma$ are useful quantities. We use subscripts to express derivatives, with actions on $x$ and $y$ distinguished by a prime. For instance, $\nabla_{y^d} \nabla_{x^c} \nabla_{y^b} \nabla_{x^a} \sigma(x,y)=\sigma_{ab'cd'}(x,y)$. In this notation (\ref{eq:mwf}) becomes $g_{ab}(x)=-\lim_{y\rightarrow x}\sigma_{ab'}(x,y)$. A similar equation is
\begin{align}\label{eq:mwf1}
g_{ab}(x)=\lim_{y\rightarrow x}\sigma_{ab}(x,y).
\end{align}
Indices can be raised and lowered by the metric. For instance, $\tensor{\sigma}{^a_b}(x,y)=\nabla_{x^b}[g^{ac}(x)\sigma_c(x,y)]$.

Differentiating (\ref{eq:swf1}) leads to
\begin{align}\label{eq:dwf}
\sigma_a(x,y)=-(\lambda_y-\lambda_x)g_{ab}(x)t^b(x).
\end{align}
Up to $-(\lambda_y-\lambda_x)$, $\sigma_a$ agrees with the tangent vector along the geodesic at $x$. This implies
\begin{align}\label{eq:wfn}
g_{ab}\sigma^a\sigma^b=(\lambda_y-\lambda_x)^2 g_{ab}t^a t^b=2\sigma.
\end{align}
The norm of $\sigma^a$ equals $\abs{2\sigma}^{1/2}=s$, the proper distance.

\section{Van Vleck-Morette determinant}\label{sec:vd}

The Van Vleck-Morette arose early on in studying the classical limit of quantum mechanics \cite{VanVleck1928TheMechanics.} and in studying the path integral transition amplitudes \cite{Morette1951OnIntegrals}. Later on it found applications to several other subjects such as the heat kernel expansion, geometrical optics, Riemannian geometry etc. \cite{Visser1993VanSpacetimes}. 

Consider a Lorentzian $d+1$-dimensional spacetime with the metric $g_{ab}$ and world function $\sigma(x,y)$.
The Van Vleck-Morette determinant is usually defined as
\begin{align}
\Delta(x,y)=(-1)^d\frac{\det[\sigma_{ab'}(x,y)]}{\sqrt{-g(x)}\sqrt{-g(y)}},
\end{align}
which is manifestly symmetric in $x$ and $y$. 
A near coincidence expansion of $\Delta$ establishes the connection between $\Delta$ and the Ricci tensor:
\begin{align}\label{eq:dtr}
\Delta(x,y)=1+\frac{1}{6} R_{ab}(x)\sigma_a(x,y)\sigma_b(x,y)+\mathcal{O}(\epsilon^3),
\end{align}
where $\epsilon$ measures the size of of a typical component of $\sigma^a$. A proof can be found in \cite{Poisson2011}.

$\Delta(x,y)$ can be re-expressed in terms of the world function and its second order derivative without reference to the metric.
When $x$ and $y$ are timelike or spacelike with proper distance $s=\abs{2\sigma(x,y)}^{1/2}$, $\Delta(x,y)$ obeys the differential equation
\begin{align}
\frac{d\Delta(s)}{ds}=(\frac{d}{s}-\theta)\Delta(s),
\end{align}
with the boundary condition $\Delta(0)=1$  \cite{Visser1993VanSpacetimes, Poisson2011}. This differential equation can be derived starting with differentiating (\ref{eq:wfn}) twice (Section 7.2 of \cite{Poisson2011}). Here $\theta(x)=\nabla_a u^a$ is the expansion of the geodesic congruence originating from $y$ with the normalized tangent vector $u^a=\sigma^a/\norm{\sigma^a}=\sigma^a/s$. One can check that
\begin{align}
    \theta(x)=\frac{\tensor{\sigma}{^a_a}(x,y)-1}{\abs{2\sigma(x,y)}^{1/2}}.
\end{align}
The solution to the differential equation can be obtained by integration
\begin{align}\label{eq:vdts}
\Delta(s)=&C s^d \exp{-\int \theta ds'},
\end{align}
where the integral is along the geodesic from $y$ to $x$, and $C$ is a constant determined by the boundary condition $\Delta(0)=1$. As both $s$ and $\theta$ can be written in terms of $\sigma$, (\ref{eq:vdts}) expresses $\Delta$ as a functional of $\sigma$ (and its second order derivative).

The lightlike case is complicated by the fact that the proper distance vanishes and is no longer an affine parameter. Fortunately in the integral over spacetime geometries $\sigma=0$ has measure zero, so we do not need to delve into the lightlike case (although there is an expression similar to (\ref{eq:vdts}) in this case \cite{Visser1993VanSpacetimes}).

\Cref{eq:vdts} offers the intuition that the Van Vleck-Morette determinant measures the curvature of spacetime in terms of the amount of focusing/defocusing of geodesic sprays \cite{Visser1993VanSpacetimes}. Without any curvature, the transverse density of geodesics would fall by $s^{-d}$ as they reach out a proper distance $s$. With curvature, the density falls instead by $\exp{-\int \theta ds'}$. The Van Vleck-Morette determinant quantifies the ratio.

The Van Vleck-Morette determinant in the form of (\ref{eq:vdts}) interestingly resembles the Wilson line expression $\mathcal{P}\exp{i\int A_a dx^a}$. The connection to gauge theories is worth exploring.

\section{Parker's magic}\label{sec:po}

The path integral representation of the Feynman propagator of equations (\ref{eq:spt}) to (\ref{eq:mac}) is reproduced here.
\begin{align}\label{eq:spta}
G(x,y)=&i\int_0^\infty \braket{x,l}{y,0} e^{-im^2 l} dl,
\\
\label{eq:wlpia}
\braket{x,l}{y,0}=&\int d[x(l')] \exp{i\int_0^l dl' [\frac{1}{4}g_{ab}\frac{dx^a}{dl'}\frac{dx^b}{dl'}-(\xi-\frac{1}{3})R(l')]}
\\
:=&\lim_{N\rightarrow \infty}\Big[\frac{1}{i}(\frac{1}{4\pi i \epsilon})^2\Big]^{N+1}\int \prod_{n=1}^N d^4 x_n [-g(x_n)]^{1/2} 
\nonumber\\
& \exp{\sum_{m=0}^N i\int_{m\epsilon}^{(m+1)\epsilon}[\frac{1}{4}g_{ab}\frac{dx^a}{dl'}\frac{dx^b}{dl'}-(\xi-\frac{1}{3})R(l')] dl' },\label{eq:maca}
\end{align}
Parker \cite{Parker1979PathSpace} found that using
\begin{align}
&\braket{x,l}{y,0}=\lim_{N\rightarrow \infty}\Big[\frac{1}{i}(\frac{1}{4\pi i \epsilon})^2\Big]^{N+1}\int \prod_{n=1}^N d^4 x_n [-g(x_n)]^{1/2} 
\nonumber\\
& \exp{\sum_{m=0}^N i\int_{m\epsilon}^{(m+1)\epsilon}[\frac{1}{4}g_{ab}\frac{dx^a}{dl'}\frac{dx^b}{dl'}-[\xi-\frac{1}{3}(1-p)]R(l')+p\ln\Delta(x_m,x_{m+1})] dl' }\label{eq:macpa}
\end{align}
instead does not affect $G(x,y)$. The magic is that for an \textit{arbitrary} constant $p$, a multiplication by $\Delta^p(x_m,x_{m+1})$ can be compensated by $\exp{-i\int_{m\epsilon}^{(m+1)\epsilon}(R(l')p/3) dl'}$. \Cref{eq:pmf1} is an another way to express the arbitrariness of $p$, with $p=a$ on one side and $p=b$ on the other. This magical freedom in $p$ allows us to trade between an $R$ term and a $\Delta$ term. 

A proof of this freedom by carrying out the path integral is given in Parker's original paper \cite{Parker1979PathSpace}. Later Bekenstein and Parker offered a short explanation (Appendix A of \cite{Bekenstein1981Path-integralSpacetimeb}), which we review here.
Consider carrying out all the integrals in (\ref{eq:macpa}) except the last
\begin{align}\label{eq:tad}
\braket{x,l}{y,0}=\int \braket{x,l}{x_N,l-\epsilon}\braket{x_N,l-\epsilon}{y,0} \sqrt{-g}d^4 x_N,
\\
\braket{x,l}{x_N,l-\epsilon}=\Delta^p(x,x_N) \exp{i\frac{\sigma(x,x_N)}{2\epsilon}-i\epsilon[\xi-\frac{1}{3}(1-p)]R},\label{eq:ipia}
\end{align}
where (\ref{eq:swf1}) is used to introduce $\sigma$. From (\ref{eq:dtr}),
\begin{align}
\Delta^p=1+\frac{1}{6}p R_{ab}\sigma_a\sigma_b+\mathcal{O}(\epsilon^3).
\end{align}
For simplicity denote the exponential in (\ref{eq:ipia}) by $E$. Then
\begin{align}
\braket{x,l}{x_N,l-\epsilon}=\{1-\frac{2}{3}p R^{ab}\epsilon^2 [\nabla_a\nabla_b-\frac{i\sigma_{ab}}{2\epsilon}-\frac{1}{2}(\xi-\frac{1}{3}+\frac{1}{3}p)(\sigma_aR_{,b}+\sigma_aR_{,b})]\}E+\mathcal{O}(\epsilon^3).
\end{align}
If $\braket{x,l}{x_N,l-\epsilon}$ was alone, $\nabla_a\nabla_b$ acts on $E$ to create a dominating $\mathcal{O}(\epsilon^{-2})$ term in the square bracket. Yet crucially, $\braket{x,l}{x_N,l-\epsilon}$ is composed with $\braket{x_N,l-\epsilon}{y,0}$ in (\ref{eq:tad}).  $E$ multiplied by $\braket{x_N,l-\epsilon}{y,0}$ generates the transition amplitude from $y$ to $x$ through $x_N$. $\nabla_a\nabla_b$ acts on this amplitude to create a $\mathcal{O}(\epsilon^0)$ term, dominated by the $\epsilon^{-1}$ term in the square bracket. Hence as $\epsilon\rightarrow 0$,
\begin{align}
\braket{x,l}{y,0}=\int [1+\frac{i}{3}\epsilon p R^{ab}\sigma_{ab} +\mathcal{O}(\epsilon^2)]E\braket{x_N,l-\epsilon}{y,0} \sqrt{-g}d^4 x_N
\end{align}
As $N\rightarrow\infty$, $x\rightarrow x_N$, and by (\ref{eq:mwf1}) $\sigma_{ab}\rightarrow g_{ab}$, so $R^{ab}\sigma_{ab}\rightarrow R$. Thus
\begin{align}\label{eq:rs}
\braket{x,l}{y,0}=\int \exp[i\frac{\sigma(x,x_N)}{2\epsilon}-i\epsilon(\xi-\frac{1}{3})R]\braket{x_N,l-\epsilon}{y,0} \sqrt{-g}d^4 x_N.
\end{align}
The $p$-dependence has dropped out, implying that $p$ is arbitrary. The reasoning can be applied to other segments, the whole path integral is independent of $p$.

For the application to gravitational path integrals, it is important to note that the result hold for arbitrary metric fields $g_{ab}$. 
One can check that all the expressions from (\ref{eq:spta}) to (\ref{eq:rs}) are meaningful, and all the steps in the derivation carry through without assuming that $g_{ab}$ solves Einstein's equations.

\section{Path integral schemes}\label{sec:pis}

The path integral for a free non-relativistic particle is usually given in a graph refinement scheme by
\begin{align*}
\braket{t_I,x_I}{t_F,x_F} =& \lim_{N \rightarrow \infty} (\prod_{i=1}^{N-1} \int dx_i) \prod_{i=1}^{N} \mu(t_i,t_{i-1}) \exp{\frac{i}{\hbar} \frac{m}{2} \frac{(x_{i}-x_{i-1})^2}{t_i-t_{i-1}}},
\\
\mu(t_i,t_{i-1})=&\frac{1}{\sqrt{2\pi i \hbar (t_i-t_{i-1})/m}},
\end{align*}
where $x_0$ and $x_N$ are fixed at the initial and final positions $x_I$ and $x_F$, and $t_0$ and $t_N$ are fixed at the initial and final times $t_I$ and $t_F$. Here we used a $1D$ temporal graph with time steps $t_0, t_1, \cdots, t_N$ to enumerate the paths, and the graph refinement is taken by going to infinitely many time steps in $\lim_{N \rightarrow \infty}$.

In a scheme that in addition sums over temporal step locations, 
\begin{align*}
\braket{t_I,x_I}{t_F,x_F} =& \lim_{N \rightarrow \infty} (\prod_{i=1}^{N-1} \int_{t_{0}}^{t_{i+1}} d t_i \int d x_i)  \mu(N,T) \exp{\frac{i}{\hbar} \sum_{i=1}^{N} \frac{m}{2} \frac{(x_{i}-x_{i-1})^2}{t_i-t_{i-1}}},
\\
\mu(N,T)=&\frac{1}{\sqrt{2\pi i \hbar T/m}}[(\frac{2\pi i \hbar T^3}{m})^{\frac{N-1}{2}} \prod_{n=1}^{N-1} C(n)]^{-1},
\end{align*}
where $T:=t_F-t_I$, $C(n):=\sqrt{\pi}\Gamma(\frac{3n}{2})/2\Gamma(\frac{3(1+n)}{2})$, 
and $\Gamma(x)$ is the gamma function. Note the difference in the measure $\mu$ from the case above. One can check (e.g., using symbolic integration of a mathematics software) that this reproduces the standard result above.

\section{An alternative expression for scalar matter coupling}\label{sec:acsm}

As mentioned in \cref{sec:s}, there is an alternative expression for the matter amplitude that extends the $l$-integrals beyond individual edges. Each matter subgraph $\gamma$ has a corresponding Feynman diagram $\gamma_F$ with edges $K\in \gamma_F$. Each edge $k\in \gamma$ belongs to an edge $K\in \gamma_F$, and we write this as $k \in K$. In this alternative expression, the $l$-integrals are along edges $K\in \gamma_F$ that can contain multiple $k\in \gamma$. In place of (\ref{eq:ma1}) and (\ref{eq:mae}), we have the matter amplitude
\begin{align}
A_M[\gamma,g]=&V[\gamma]\prod_{K\in \gamma_F}A_M[K,g],
\\
A_M[K,g]=&i \int_0^\infty d l~ \prod_{k\in K} \int_0^\infty \frac{dl_k}{i(4\pi i l_k)^{2}} \exp{i\frac{\sigma_k}{2l_k}-i(\xi-\frac{1}{3})R_k l_k-im^2l_k} \delta(\sum_k l_k-l).
\label{eq:x1}
\end{align}
The $i$ up front in (\ref{eq:x1}) comes from the $i$ up front in (\ref{eq:spt}). The individual $i$ beneath $dl_k$ comes from the same factor in (\ref{eq:mac}). These $i$ factors cancel each other in (\ref{eq:mae}) when we perform an independent $l$ integral on each edge, but they stay here. One could use the integral representation $\delta(x)=\int_{-\infty}^\infty  \frac{dy}{2\pi} e^{-ixy}$ to rewrite (\ref{eq:x1}) as
\begin{align}
&A_M[K,g]
\nonumber
\\
=&i\int_0^\infty d l~  \prod_{k\in K} \int_0^\infty \frac{dl_k}{i(4\pi i l_k)^{2}} \exp{i\frac{\sigma_k}{2l_k}-i(\xi-\frac{1}{3})R_k l_k-im^2l_k} \int_{-\infty}^\infty  \frac{dy}{2\pi} e^{-i(\sum_k l_k-l)y} 
\\
=&i\int_{-\infty}^\infty  \frac{dy}{2\pi} \prod_{k\in K} \int_0^\infty \frac{dl_k}{i(4\pi i l_k)^{2}} \exp{i\frac{\sigma_k}{2l_k}-i(\xi-\frac{1}{3})R_k l_k-i(m^2+y)l_k} \int_0^\infty d l ~ e^{ily}
\\
=&\int_{-\infty}^\infty  \frac{dy}{-2\pi y} \prod_{k\in K} \int_0^\infty \frac{dl_k}{i(4\pi i l_k)^{2}} \exp{i\frac{\sigma_k}{2l_k}-i(\xi-\frac{1}{3})R_k l_k-i(m^2+y)l_k}.
\end{align}
In the last step we performed the $l$ integral, assuming that $y$ has an infinitesimal imaginary part to make the integral converge.

Given $\Gamma$ as a spacetime graph, and $\gamma_\Gamma$ as a matter subgraph with corresponding Feynman diagram $\gamma_F$, we combine $A_M[\gamma_\Gamma,g]$ with the gravitational edge amplitudes to obtain
\begin{align}
A[\Gamma,\gamma_\Gamma,g] =& V[\gamma_\Gamma]\prod_{\kappa\subset \Gamma}\prod_{h\in\kappa} \frac{\sigma_h}{\sigma_h-6i \eta_{\kappa,h} \Delta_h^{-1} \ln \Delta_h} 
\nonumber
\\
&\prod_{K\in \gamma_F} \int_{-\infty}^\infty  \frac{dy}{-2\pi y} \prod_{k\in K}\Delta_k^{1-3\xi} \int_0^\infty \frac{dl_k}{i(4\pi i l_k)^{2}}
\exp{i\frac{\sigma_k}{2l_k}-i(m^2+y)l_k},
\label{eq:gmab2}
\end{align}
where as in the pure gravity case we applied (\ref{eq:pmf2}) to the matter amplitude to trade away the Ricci curvature term for a $\Delta$ term. Finally (\ref{eq:gmab2}) should be plugged into (\ref{eq:gma1}) and (\ref{eq:gma2}) to obtain the gravity-matter amplitude $A_{GM}$.


\section{Comparison with related approaches}\label{sec:cra}

There are other path integral approaches of quantum gravity based on graphs and dynamical variables assigned to the edges. 
We discuss connections to and differences from some of these approaches.

\textbf{Quantum Regge Calculus.} Quantum Regge Calculus \cite{Loll1998DiscreteDimensions, Hamber2009QuantumApproach, Barrett2019TullioGravity} is similar in using the squared invariant distance $\sigma$ as a basic variable. 

Some differences are: 1) Quantum Regge Calculus captures gravity geometrically, while WQG captures gravity correlationally. Quantum Regge Calculus is based on piecewise flat simplicial geometries (which induce triangle inequality constraints), and captures gravity geometrically by curvatures measured by deficit angles on the hinges of the simplicies. WQG does not assume piecewise flatness for the lattice units, and captures gravity by its effects on matter correlations mediated by quantum test fields/particles. 2) Quantum Regge Calculus is mostly studied in the Euclidean signature (see 
\cite{Williams1986QuantumFormulation, Sorkin1974DevelopmentFields, *Sorkin1975Time-evolutionCalculus, *SorkinLorentzianVectors, Tate2011Fixed-topologyDomain, *Tate2012Realizability1-simplex, AsanteEffectiveGravity}  for a few exceptions). WQG is directly formulated in the Lorentzian regime with an motivation for studying superpositions of spacetime causal structures.

\textbf{Dynamical Triangulations.} Dynamical triangulation models \cite{Weingarten1982EuclideanLattice, Loll1998DiscreteDimensions, Ambjrn2001DynamicallyGravity, Ambjorn2012NonperturbativeGravity} make use of Monte Carlo simulation to carry out quantitative studies. The same is expected for WQG.


Some differences are: 1) Like Quantum Regge Calculus, dynamical triangulation models also make use of the Regge action. Piecewise flat geometry is assumed, in contrast to WQG. 2) In the Lorentzian models of Causal Dynamical Triangulation \cite{Ambjrn2001DynamicallyGravity}, a global time foliation is assumed. Most studies inherit this assumption, although models dropping it have been discussed \cite{Jordan2013CausalFoliation}. The WQG models presented here do not assume a global time foliation. 

\textbf{Spin-foam Models and Group Field Theories.} Spin-foam Models and Group Field Theories \cite{Ponzano1968SemiclassicalCoefficients, ReisenbergerWorldsheetGravity, Baez1997SpinModels, Perez2013TheGravity, Rovelli2014CovariantGravity, Freidel2005GroupOverview} are similarly path integral approaches based on graphs. Group variables are central to these approaches. At the end of \cref{sec:vd} we noted that the resemblance between the Van Vleck-Morette determinant and the Wilson line. It would be interesting to study if there is a gauge theory aspect of the WQG approach. 


For Spin-foam Models and Group Field Theories, there is no consensus among the researchers whether the graphs are embedded in continuum spacetime manifolds or are standalone fundamental structures. Nor is there consensus whether the graphs should be summed over.
For WQG the graphs are assumed to be standalone fundamental structures, as mentioned in \cref{sec:ts}. Both the graph refining and graph summing schemes are considered, and they could lead to the same results for suitable path integral measures, as mentioned in \cref{sec:pi}.

For Spin-foam Models and Group Field Theories, the edges of the graphs are usually taken to have a fixed signature (timelike in most models) \cite{BahrOnModels}. It is unclear to what extent spacetimes are summed over beyond conformal degrees of freedom. For WQG the edges do not have a fixed signature, and causal structures are summed over.


\textbf{Causal Sets.} The Causal Set approach \cite{bombelli_space-time_1987, Benincasa2010ScalarSet, Surya2019TheGravity} is similar in emphasizing the importance of spacetime causal structures. In a causal set the causal relation is given by a partial ordering. In WQG the causal relation is given by the sign of $\sigma$. The two approaches are also similar in treating continuum spacetime manifold as non-fundamental.

The Causal Set approach assumes that spacetime is fundamentally discrete. This assumption is needed to capture the conformal factor information of spacetime configurations with the number of points. WQG does not assume fundamental discreteness of spacetime from the outset. Continuous values of $\sigma$ are summed over, including zero. The vision is that an ultraviolet regularization is generated dynamically when different causal relations are summed over in the path integral  \cite{JiaUltravioletStructure}.

\bibliographystyle{apsrev4-1}
\bibliography{mendeley.bib}

\end{document}